\documentclass[onecolumn,showpacs,preprintnumbers,amsmath,amssymb]{revtex4}

\usepackage{epsf}
\usepackage{graphicx}  
\usepackage{dcolumn}   
\usepackage{bm}        

\newcommand{\be}{\begin{equation}}
\newcommand{\en}{\end{equation}}
\newcommand{\bea}{\begin{eqnarray}}
\newcommand{\ena}{\end{eqnarray}}
\newcommand{\s}{\sqrt}
\newcommand{\ar}{{\arctan}}
\newcommand{\n}{\nonumber}
\newcommand{\G}{\Gamma}
\begin{document}


\title{A new class of plane symmetric solution}

 \author{Hongsheng Zhang\footnote{Electronic address: hongsheng@kasi.re.kr} }
 \affiliation{\footnotesize
 Korea Astronomy and Space Science Institute, Daejeon 305-348, Korea }
 \affiliation{\footnotesize Department of Astronomy, Beijing Normal University,
Beijing 100875, China}
 \author{Hyerim Noh\footnote{Electronic address: hr@kasi.re.kr} }
 \affiliation{\footnotesize
 Korea Astronomy and Space Science Institute,
  Daejeon 305-348, Korea }
 \author{Zong-Hong Zhu\footnote{Electronic address:
zhuzh@bnu.edu.cn}}
  \affiliation{\footnotesize Department of Astronomy, Beijing Normal University,
  Beijing 100875, China}
 \date{ \today}

\begin{abstract}
 A new class of static plane symmetric solution of Einstein field equation generated by a perfect fluid source
   is put forward. A special family of this new solution is investigated in detail.
    The constraints on the parameters by different energy conditions are
 studied. The classical stability of this solution is discussed.
  The junction conditions matching to Minkowski metric and Taub metric
 are analyzed respectively.

\end{abstract}


\maketitle

\section{Introduction}
 Though considerable amounts of vacuum solutions of Einstein equation are known,
 but, frankly speaking, the physical
interpretation of many of them remains unsettled
\cite{exactsolution}. As emphasized in \cite{bonnor}, the key to
physical interpretation is to find out the nature of the sources
which generate these vacuum spaces.  Singularity theorem is one of
the most important and profound theorem in classical general
relativity. For a deeper understanding and control of the
singularities, the physical characters, such as mass densities,
internal pressures, of the sources are imperative. For example in
black hole solutions, where only singularity ``generates'' the whole
space, investigating of how a matter distribution gives rise to the
black hole space is necessary to judge the physical significance (or
lack of it) of the complete analytic extensions of these solutions.

Concerning how we obtain a comprehending of the sources, there is
really no alternative  for constructing an interior solution for a
matter distribution which matches to the vacuum space in question.
We will present a new class of plane symmetric solution sourced  by
perfect fluid. One may be curious about where we can find a
realistic plane symmetric space. In the researches of the early
universe, people find there may be several-time phase transitions
when the universe
 is cooled down, and many topological defects, such as monopole,
 cosmic string, and domain wall may survive today. Theoretically, equivalence principle, as
 principle of general relativity,  says that a gravity field can be
 simulated by an accelerating effect. For the case of a uniformly
 accelerating observer, an exact simulation should be plane
 symmetric (but curved) space.

  The static vacuum plane symmetric solution, Taub solution was
  obtained more than 50 years ago  \cite{taub}. The solution is
  asymptotically flat but there is a serious time-like singularity in Taub space.
  People believe that the singularity should be superseded by some
  matter source, for some attempts to seek the sources, see \cite{source}.
   Taub solution permits a
  three-dimensional Euclidean symmetry group, but it is not homogeneous. Here a solution is called
  homogeneous if it permits three  translational space-like Killing vectors. Taub solution
  only permits two: It is not homogeneous in $z$ direction. Further, it is shown that generally speaking
  any singularity free source with reflective symmetry for plane symmetric vacuum space
  must violate dominant energy condition \cite{dec}.

   In electromagnetics  the electronic field of
  an infinite homogeneously charged flat board is homogeneous.  By
  analogy, in linearized gravity theory the solution for a $\delta$-function
  like source has been derived \cite{delta}, which is
  spatially homogeneous and seems to contradict with \cite{dec}.
   It is suggested  that the counterpart of the Einstein theory to the
  gravitational field of a massive Newtonian plane should be described
  by the Rindler solution rather than Taub's \cite{wang}.

  We will present a new class of plane symmetric solution with
  perfect fluid source, which definitely violates the dominant energy
  condition. We study some various properties of this new solution
  and do some preliminary studies on how to glue it to a vacuum space.

 This paper is organized as follow: In the next section we will
 present our solution. Because
 the energy momentum tensor is very miry and hence the physical properties
 of this solution are difficult to analyse,  the full
 form of the energy momentum tensor is casted into the appendix.
 We then give a special
 class of the solution as a starting point of the further
 discussions. In section III we study the constraint on the parameters by
  the different energy conditions.  In section IV, we investigate the stability property of this solution.
 In section V, we try to glue this sourced solution to vacuum solutions. The cases of Minkowski and Taub are
 discussed respectively.  A
    summary and some discussions are presented in section VI.

 \section{the solution}
  We find that the following plane symmetric metric
  \bea
 ds^2=-f(z)^2dt^2+dz^2+
 e^{2az}f(z)^2e^{-2\left[az+\frac{h(z)\arctan\left(e^{az}
 \sqrt{\frac{w}{c}}\right)}{\sqrt{wc}}-dh(z)\right]}\left(dx^2+dy^2\right),
 \label{general}
 \ena
 solves the Einstein field equation with a perfect fluid source.
 Here $a,~w,~c,~d$ are 4 real parameters, and
 \be
 f(z)=ce^{-az}+we^{az},
 \en
 \be
 h(z)=-ce^{-az}+we^{az}.
 \en

 Clearly there are four Killing fields $\frac{\partial}{\partial t}$,
  $\frac{\partial}{\partial
 x}$,  $\frac{\partial}{\partial y}$ and $-y\frac{\partial}{\partial x}+x \frac{\partial}{\partial y}$. The latter three  span a Euclidean group $G_3$,
 which implies the plane symmetry.
 The metric (\ref{general}) solves the Einstein field
equation with a source in  perfect fluid form
 \be
 T=(\rho(z)+p(z))U\otimes U+p(z)g,
 \label{em}
 \en
 where $T$ denotes the energy momentum tensor of the fluid, $U$
 stands for four-velocity of the fluid and $g$ denotes the metric
 tensor.
 Since the exact forms of $p(z)$ and $\rho(z)$ are very involved,
  we just present them in the appendix. The energy  momentum
 tensor of metric (\ref{general}) is too complicated to make further
 analysis. Alternatively we may concentrate on a special family of solution
 (\ref{general}).

  First, we set $w=0$, and hence (\ref{general}) reduces to
 \bea
 ds^2=-c^2e^{-2az}dt^2+dz^2+
 c^2e^{-2(az-1+dce^{-az})}(dx^2+dy^2),
 \label{general1}
 \ena
 where we have used
 \be
 \lim_{w\to 0} \frac{h(z)\arctan\left(e^{az}
 \sqrt{\frac{w}{c}}\right)}{\sqrt{wc}}=-1.
 \en
 Then, we rescale the coordinates $t$, $x$ and $y$ by
 \be
 ct \to t,~~~~~~~~ecx \to x,~~~~~~~~ecy\to y.
 \en
 So (\ref{general1}) further reduces to
 \bea
 ds^2=-e^{-2az}dt^2+dz^2+e^{-2(az+de^{-az})}(dx^2+dy^2).
 \label{general2}
 \ena
 Introducing new variables
 \be
 a'=-a,~~~~~~~b'=-d,
 \en
 and for convenience labeling $a'$ by $a$ and $b'$ by $b$, we derive
 a special solution as follow,
 \be
 ds^2=-e^{2az}dt^2+dz^2+e^{2(az+be^{az})}(dx^2+dy^2).
  \label{simmet}
 \en
 There are  two free parameters in the above solution.
 With the above metric $\rho(z)$ and $p(z)$, respectively given by
 (\ref{density}), (\ref{pressure}) for the original metric (\ref{general}),
 are reduced to extraordinarily simple form
 \bea
 \rho(z)&=&-a^2(3+8be^{az}+3b^2e^{2az}),
 \label{rho} \\
 p(z)&=&a^2(3+4be^{az}+b^2e^{2az}),
 \ena
 where we have set $8\pi G\equiv 1$, and $G$, as usual, denotes the Newton gravitational constant. In this article we
 will concentrate on metric (\ref{simmet}).
      Before studying the detailed properties of (\ref{simmet}), we
 first investigate the various limitations of it. Obviously when
 $a=0$ (\ref{simmet}) degenerates to Minkowski metric. When $b=0$, it
 becomes
 \be
 ds^2=dz^2+e^{2az}(-dt^2+d{x}^2+d{y}^2),
 \en
  which is just anti-de Sitter (AdS)
 metric in an unattractive coordinate system.
 Surely it describes a homogeneous and isotropic solution in
 all directions, not only a plane symmetric one.

 \section{Energy conditions}
   More and more cosmological evidences
  imply that exotic matters violating energy conditions, such as weak,
  strong, dominant or null energy conditions, do exist in our
  universe. And the  partition of the exotic matter takes a much
  more value than that of dust matter.
   But different requirements on the energy-momentum tensor have been used to prove several pivotal theorems, for example
  singularity theorem and positive energy theorem,  in
  classical general relativity. It may be
   helpful to investigate the properties of the matter in the source by different
   energy conditions. Further, it is pointed out that the source of Taub must violate the dominant energy condition
    if the source is reflective symmetric and without spacetime singularity \cite{dec}.  Now we begin to
   study the parameter regions permitted by weak, strong and dominant energy
   conditions. Note that weak and strong energy conditions are two
   independent conditions. In this section we consider a plane
   source with finite thickness, which resides in the region $z\geq
   0$.
   Our solution is an interior solution of plane source, thus the pressure
   $p(z)$ should vanish naturally at some distance from the ``ground",
   $z=z_0 $, that is, we should match it to a vacuum solution. This
   condition implies

 \be
 p(z_0)=a^2(3+4be^{az_0}+b^2e^{2az_0})=0.
 \en
 The two roots of the above equation reads
 \bea
 b_1=-e^{-az_0},\\
 b_2=-3e^{-az_0}.
 \ena
 $b_1$ and $b_2$ denote the two branches of the solution which
 have finite thickness. We call them ``little branch" and ``large
 branch", respectively.

   To show the physical meanings of the parameters, we calculate the
   mass per area of this slab $\alpha$.
  \be
  \alpha=\int_0^{z_0}  dz e^{2\left[az+be^{az}\right]} \rho(z),
  \en
  where $\rho(z)$ is defined in (\ref{rho}). For the little branch $b=-e^{-az_0}$,
  the integral yields,
  \bea
  \alpha=\frac{a}{8}e^{-az_0}\left[-5e^{-2+3az_0}+e^{-2e^{-az_0}}
  \left(14e^{az_0}+2e^{2az_0}+e^{3az_0}-12\right)\right].
  \label{aremas1}
  \ena
  For the large branch, the result is
 \bea
  \alpha=\frac{a}{72}e^{-az_0}\left[191e^{-6+3az_0}+e^{-2e^{-az_0}}
  \left(126e^{az_0}+6e^{2az_0}+e^{3az_0}-324\right)\right].
  \label{aremas2}
  \ena
 We see that the  mass per area is proportional to $a$, so it
 can be treated as a mass parameter of the slab.
  Whether the stuff filled in the slab is
  ordinary or exotic in some sense depends on the value of $az_0$. Different energy conditions impose different
  constraints on $az_0$.

 Weak energy condition (WEC) requires that the energy momentum
 tensor of the source $T$ satisfies,
 \be T(Z,Z)\geq 0, \en
 for any time-like vector $Z$, which equals
 \bea
 \rho=-a^2(3+8be^{az}+3b^2e^{2az})\geq 0,
 \label{1stwec} \\
 \rho+p=-a^2(4be^{az}+2b^2e^{2az})\geq 0
 \label{2ndwec}.
 \ena

 First, we consider the little branch $-e^{-az_0}$. We then replace $b$ by $-e^{-az_0}$
 in (\ref{1stwec}) and (\ref{2ndwec}). For further discussions we
 investigate three cases depending on the sign of $a$.

 Case I:
 $a>0.$
  Inequality
 (\ref{1stwec}) requires
 \be
 e^{az_0}\leq\frac{3}{4-\sqrt{7}},
 \en
 and inequality (\ref{2ndwec}) is satisfied spontaneously in
 this settlement.

 Case II:
 $a<0.$ Inequality
 (\ref{1stwec}) requires
 \be
 e^{az_0}\geq \frac{3}{4+\sqrt{7}},
 \en
 Inequality
 (\ref{2ndwec}) requires
 \be
 e^{az_0}\geq 1/2,
 \en

 therefore the permitted range of $a$ and $z_0$ is
 \be
 e^{az_0}\geq 1/2.
 \en
 The case $a=0$ degenerates to Minkowski space, which is a trivial
 case  satisfying any energy conditions. In the interval,
  \be
  e^{az_0}\in[1/2, ~\frac{3}{4-\sqrt{7}}],
  \label{wec}
  \en
  WEC always can be satisfied for any real $a$.

  Second, we consider the large branch $b=-3e^{-az_0}$. Mimicking the above
  discussions for the case of little branch we find that WEC can not be satisfied for any $a$ and $z_0$.

 Then we turn to the strong energy condition (SEC), which requires
 \be
 Ric(Z,Z)\geq 0,
 \label{str}
 \en
 where $Ric$ denotes the Ricci tensor of metric (\ref{simmet}), and
 $Z$ is an arbitrary time-like vector. The condition (\ref{str})
 equals to
 \bea
  \rho+p=-a^2(4be^{az}+2b^2e^{2az})\geq 0,
 \label{1ststr} \\
 \rho+3p=a^2(6+4be^{az})\geq 0.
 \label{2ndstr}
 \ena

 First, similarly, we consider the little branch $b=-e^{-az_0}$. Also, we investigate
 the three cases depending on the sign of $a$, respectively.

 Case I:
 $a>0$. The inequality (\ref{1ststr}) has been discussed in the
 case WEC, which is satisfied naturally. One finds (\ref{2ndstr})
 is also satisfied naturally. So SEC imposes no constraint on the
 parameters in this case.

 Case II:
 $a<0$. Inequality
 (\ref{1ststr}) has been analyzed before, which requires
 \be
 e^{az_0}\geq 1/2,
 \en
 and inequality
 (\ref{2ndstr}) requires
 \be
 e^{az_0}\geq 2/3.
 \en
 The case $a=0$ degenerates to Minkowski space, as we have  pointed out,
  which is a trivial case marginally satisfying any energy conditions.
  For arbitrary $a$ the requirement of SEC always can be satisfied in the
  interval,
  \be
  e^{az_0}\in[2/3, \infty).
  \label{sec}
  \en

  Second, we consider the branch $b=-3e^{-az_0}$ for SEC. Mimicking the above
  discussions we find that SEC can not be satisfied for any $a$ and $z_0$.
 We see the case $b=-3e^{-az_0}$ may violate both WEC and SEC for
 any values of parameters $a$ and $z_0$.

 The dominant energy condition (DEC) requires
 \be
 \rho\geq |p|.
 \en
 For removing the calculation of absolute value, we consider the cases of $p<0$ and $p\geq 0$, respectively.
 First, $p<0$ requires that
 \be
 (b+e^{-az_0})(b+3e^{-az_0})<0,
 \en
 whose solution is an empty set.
 Second, for the case $p\geq 0$, DEC requires
 \be
  3+6be^{az}+2b^2e^{2az}\leq 0,
 \en
 which does not have non-empty solution set either. Hence we conclude that this solution always violates DEC.
  Recalling
 that \cite{dec} demonstrates that any reasonable source of Taub solution must violate
 DEC, one sees that our present solution just satisfies this requirement.

 \section{stability}
 In this section we give an initiative discussion on the global stability of
 the solution. The basic idea of our investigation is best presented
 by considering  an analogy. A Newtonian celestial body, for
 instance, the Globe, is stable, which can be understood as follow: If the Globe undergoes an small
 perturbation, expanding or contracting a little, the total energy (including
 gravitational energy and the energy of the particles of which the Globe consists) of new
 configuration is higher than the original configuration. Hence the
 Globe is stable in view of least energy principle.

  A similar
 argument can be applied to our solution. The first problem is the
 energy of the gravitational field. People recognized that there is
 no covariant energy momentum of gravitational field long ago.
 Different expressions correspond to different boundary conditions
 in the Hamiltonian approach of quasi-local mass \cite{chen}. Here
 we adopt the Landau-Lifschitz expression \cite{landau}.
 According to Landau-Lifschitz expression, the energy momentum of
 gravitational field reads
 \bea
 S^{\mu \nu}=\frac{1}{2}\left[(2\G^{\alpha}_{\beta
 \gamma}\G^{\rho}_{\alpha \rho}- \G^{\alpha}_{\beta
 \rho}\G^{\rho}_{\gamma
 \alpha}-\G^\alpha_{\beta\alpha}\G^\rho_{\gamma \rho}) (g^{\mu
 \beta} g^{\nu \gamma}-g^{\mu\nu}g^{\beta\gamma}) +g^{\mu\beta}
 g^{\gamma\alpha}(\G^\nu_{\beta\rho}\G^\rho_{\gamma\alpha}
 +\G^\nu_{\gamma\alpha}\G^\rho_{\beta\rho}-\G^\nu_{\alpha\rho}
 \G^\rho_{\beta\gamma}-\G^\nu_{\beta\gamma}\G^\rho_{\alpha\rho})\right. \n
 \\
 +\left. g^{\nu\beta}g^{\gamma\alpha}(\G^\mu_{\beta\rho}\G^\rho_{\gamma\alpha}
 +\G^\mu_{\gamma\alpha}\G^\rho_{\beta\rho}-\G^\mu_{\alpha\rho}
 \G^\rho_{\beta\gamma}-\G^\mu_{\beta\gamma}\G^\rho_{\alpha\rho})
 +g^{\beta\gamma}g^{\alpha\rho}(\G^\mu_{\beta\alpha}
 \G^\nu_{\gamma\rho}-\G^\mu_{\beta\gamma}\G^\nu_{\alpha\rho})\right],~~~~~~~~~~~~~~~~~~~~~~~~
  \label{smu}
  \ena
 where $\G$ denotes the affine connection of metric (\ref{simmet})
 and $g_{\mu\nu}$ denotes the component of (\ref{simmet}). The
 non-zero components of $\G$ are
 \bea
 \G^t_{tz}=\G^t_{zt}=a,~~~
 \G^z_{tt}=ae^{2az},~~~\G^z_{xx}=-ag_{xx},~~~\G^z_{yy}=-ag_{xx},\n
 \\
 ~~~\G^x_{zx}=\G^x_{x z}=a(1+be^{az}),~~~
 \G^y_{yz}=\G^y_{zy}=a(1+be^{az}).
 \label{gamma}
 \ena
 Substituting (\ref{gamma}) into (\ref{smu}) we derive the energy
 density of gravity,
 \be
 \rho_{\rm gra}=-S^0_0={a^2}(4+4be^{az}).
 \en

 Then, in the little branch the density per area of gravity is
 \be
 \beta=\int_0^{z_0} dz e^{2\left[az-e^{a(z-z_0)}\right]} \rho_{\rm
 gra}.
 \label{grad}
 \en
  Performing the calculation directly, we obtain
  \footnote{However, we should keep alert on this result since it is
 well known
 uncertainty of the gravitational energy density. The
 quasi-local energy defined in \cite{brown} equals the value of the
 Hamiltonian that generates unit time translations orthogonal to a
 spacelike hypersurface $D$ at its boundary two-surface $\partial D$.
 It seems a reasonable definition of the quasi-local energy for
 gravity because of this property. Now, in our case, $D$ is a
 unit-square thick slab, and naturally, $\partial D$ is
 composed of the four walls and two covers of it. The
 3-boundary is defined as the product of $\partial D$ with timelike
 world lines orthogonal to $D$ at $\partial D$. The induced 3-metric
 on the 3-boundary can be written as
 $$\gamma=-e^{2az}dt^2+dz^2+e^{2(az+be^{az})}dx^2,$$
 on the walls along $x$ direction,
 $$\gamma=-e^{2az}dt^2+dz^2+e^{2(az+be^{az})}dy^2,$$
 on the walls along $y$ direction,
  and
 $$\gamma=-e^{2az}dt^2+e^{2(az+be^{az})}(dx^2+dy^2),$$
 on the two covers, respectively. Note that the induced metric
 is not well defined on the joint lines gluing the slab
 and the two covers and the jointing lines between the walls. Hereby the boundary energy momentum tensor is
 not well defined there. Just as well the six lines are zero-measure
 sets, and then we can omit the energy associated with it without changing the result.
  With this construction it is easy
 to show that the quasi-local gravitational energy enclosed by the slab and
 is exactly the same as the result by using Landau-Lifschitz expression for our settlement.}
 \be
 \beta={2a}\left(e^{-2+2az_0}-e^{-2e^{-az_0}}\right),
 \label{gradenl}
 \en
 which is also proportional to $a$.
 For the large branch, the area density becomes
 \be
 \beta=\int_0^{z_0}  dz e^{2\left[az-3e^{a(z-z_0)}\right]} \rho_{\rm
 gra},
 \en
 which yields,
 \be
 \beta={2a}\left(e^{-6+2az_0}-e^{-6e^{-az_0}}\right).
 \label{gradenb}
 \en

  The total mass associated with the unit slab is
 $\sigma=\alpha+\beta$. In the little branch, by using (\ref{aremas1}) and (\ref{gradenl}),
 \bea
 \sigma &=& \alpha+\beta \n \\
  &=& \frac{a}{8}e^{-az_0}\left[11e^{-2+3az_0}+e^{-2e^{-az_0}}
 \left(-2e^{az_0}+2e^{2az_0}+e^{3az_0}-12\right)\right].
 \label{totalmlit}
 \ena
 Imitating the case of a Newtonian star, we impose a perturbation
  to $z_0$,
  \be
  \delta \sigma=\frac{a^2}{4} e^{-2-2e^{-az_0}-2az_0}\left(
  -12e^2+4e^{2+az_0}+2e^{2+2az_0}+2e^{2+3az_0}+e^{2+4az_0}+11e^{2e^{-az_0}+4az_0}
  \right) \delta z_0.
  \en
 Numerical calculations show that stagnation points dwell at
 $z_0=-0.292a^{-1}$ and $z_0=-1.245a^{-1}$. Further we find
 \bea
 \delta^2\sigma |_{r_0=-0.292a^{-1}}<0,
 \\
 \delta^2\sigma |_{r_0=-1.245a^{-1}}>0,
 \ena
  where we have used $a<0$, since $z_0>0$.
 We conclude that the configuration described by (\ref{simmet}) is
 stable when the parameters $a,~~r_0$ are constrained by
 $ar_0=-1.245$. The resulting stable metric is a one parameter family,
 \be
 ds^2_{\rm sta}=-e^{2az}dt^2+dz^2+e^{2(az-3.473e^{az})}(dx^2+dy^2).
 \en

  In the large branch, similar to the little branch, the total mass
  of a unit slab reads,
  \bea
  \sigma &=& \alpha+\beta \n \\
   &=& \frac{a}{72}e^{-az_0}\left[335e^{-6+3az_0}+e^{-6e^{-az_0}}
  \left(-18e^{az_0}+6e^{2az_0}+e^{3az_0}-324\right)\right].
 \label{totalmbig}
  \ena
   Imposing a perturbation to $z_0$, we reach
   \be
   \delta \sigma=\frac{a^2}{36} e^{-6-6e^{-az_0}-2az_0}\left(
  -972e^6+108e^{6+az_0}+18e^{6+2az_0}+6e^{6+3az_0}+e^{6+4az_0}+335e^{6e^{-az_0}+4az_0}
  \right) \delta z_0.
   \en

   Numerical calculations show that two stagnation points for $\delta \sigma=0$
    inhabit at $z_0=-0.3163a^{-1}$ and $z_0=0.9770a^{-1}$. Further,
    one can show
 \bea
 \delta^2\sigma |_{r_0=-0.3163a^{-1}}>0,
 \\
 \delta^2\sigma |_{r_0=0.9770a^{-1}}>0,
 \ena
 therefore, both of the two stagnation points are stable. They
 correspond to two spaces of one parameter family,
  \be
 ds^2_{\rm sta1}=-e^{2az}dt^2+dz^2+e^{2(az-4.116e^{az})}(dx^2+dy^2),
 \en
  \be
 ds^2_{\rm sta2}=-e^{2az}dt^2+dz^2+e^{2(az-1.13e^{az})}(dx^2+dy^2),
 \en
 respectively.


 \section{matching to vacuum solutions}

 Surely, (\ref{general}) and (\ref{simmet}) are rigorous solutions
 with perfect fluid sources, which can be filled in the whole space.
 However, there is a true singularity if interval of $z$ is not confined.
 In fact, the Ricci scalar $R$ reads,
  \be
  R=-2a^2(6+10be^{az}+3b^2e^{2az}).
  \en
 When $z$ goes to $\infty$ ($-\infty$), Ricci scalar will be divergent for
 a positive (negative) $a$. We hereby consider the case that this
 solution is only valid in a finite region and the spacetime is
 vacuum out of this region.

 The gravitational field must satisfy
 two boundary conditions: 1. The metric is continuous across the
 boundary surface, and 2. The extrinsic curvatures measured by the
 different sides of the boundary surface relate to each other by
 \be
 [K-h{\rm tr}(K)]^{\pm}=- \tau,
 \label{jump}
 \en
 in which $K$ denotes the extrinsic curvature of the boundary,
 $h=g-\frac{\partial}{\partial z}\otimes \frac{\partial}{\partial z}$
  represents the induced metric on the boundary,
 $\tau$ is the energy-momentum tensor confined to the boundary, and
 $[~]^{\pm}$ denotes the jump at the boundary, i.e., for a quantity
 $Q$, $[Q]^{\pm}=\lim_{{(z-z_0)} \to 0^{+}}Q(z)-\lim_{(z-z_0) \to
 0^{-}}Q(z)$.
 The exterior vacuum space has to be plane symmetric to match to
 metric (\ref{simmet}).  First we study the most simple case that the vacuum out of the source region is Minkowskian
  geometry. Then we analyse the junction condition matching to the well known
  non-flat plane symmetric space: (static) Taub space.

  Before discussing the boundary condition between the slab and the vacuum, we
  impose a $Z_2$ (reflective) boundary condition at $z=0$. The continuity condition is naturally
  satisfied and the jump condition generates
  \be
  [K-h{\rm tr}(K)]^{0+}=-\frac{1}{2} (\tau) ^{0},
 \en
 where $0+$ labels the value of a quantity at $z=0$  going
  from the positive direction, and $0$ denotes the value of a quantity
  at $z=0$. Using the above equation we derive
  \be
  \tau^{\nu}_{\mu}=2 {\rm diag}(2a(1+b),a(2+b),a(2+b)),
  \en
  where $b=-e^{-az_0}$ or $b=-3e^{-az_0}$, depending on the little
  or large branch.
      Now we write the Minkowskian metric in the
  following chart,
  \be
  ds^2_{Min}=-m^2dt^2+dz^2+n^2(dx^2+dy^2),
  \label{min}
  \en
 where $m$ and $n$ are positive constants. The continuous condition yields
 \bea
 m&=&e^{az_0},\\
 n&=&e^{az_0+be^{az_0}},
 \ena
 which corresponds to ``time contraction'' and ``length contraction'',
 though $m,~n,$  may be greater than 1 so
 that they represent ``time dilation'' and ``length dilation''.
   From the jump condition (\ref{jump}) we obtain
 $\tau$ in induced chart by (\ref{min}) in the little branch,
 \be
 \tau_{~\mu}^{\nu}={\rm diag}({0,~-a,-a}),
 \en
 while in the large branch,
 \be
 \tau_{~\mu}^{\nu}={\rm diag}({4a,~a,~a}).
 \en
  $\tau$ vanishes when $a=0$, which is consistent with our expectation because
 the interior metric degenerates to Minkowski when $a=0$.

 Next we discuss the conditions matching to the Taub metric. The Taub's metric
 reads,
 \be
 ds^2=-z^{-2/3}k^2dt^2+dz^2
 + z^{4/3}l^2\left(dx^2+dy^2\right),
 \label{taub}
 \en
 where $k, l$ are two positive constants.
 The continuous condition requires
 \bea
 3az_0=1+2\ln kl,
 \label{azl}
 \ena
 in the little branch, and
 \be
 az_0=1+\frac{2}{3}\ln kl,
 \label{azb}
 \en
 in the large branch.

   The jump condition (\ref{jump}) gives, in
 the induced chart by (\ref{taub}) for the little branch
 \be
 \tau_{~\mu}^{\nu}={\rm diag}\left(-\frac{8}{3}z_0^{-5/3},~\frac{2}{3}z_0^{-1}+a-\frac{4}{3}z_0^{-5/3},~
 \frac{2}{3}z_0^{-1}+a-\frac{4}{3}z_0^{-5/3}\right),
 \label{taulit}
 \en
 where $z_0$ is given by (\ref{azl}). And for the large branch

 \be
 \tau_{~\mu}^{\nu}=(a+\frac{1}{3}z_0^{-1}){\rm
 diag}\left(-4,~-1,-1\right),
  \en
 where $z_0$ takes the value in (\ref{azb}). If we require the
 matching is perfect, that is, boundary energy momentum vanishes, we
 arrive at $az_0=-1/3$.

  On the other
 hand, if perfect matching condition is imposed in the little
 branch, one needs an infinitely thick slab, which can be derived
 from (\ref{taulit}).

\section{Conclusions and discussions}
 In this article we present a global plane symmetric solution of Einstein field equation
  with a perfect fluid source, which we interpret as the source of some plane symmetric vacuum space.
 Four Killing vectors, including a time like Killing vector,
 are permitted in this solution. We find a chart in which the metric is
 written in time coordinate orthogonal form. Then we give a
 special and much simpler class of the original solution as the starting point for
 further investigations.

 We find the ranges of the parameters in which WEC and SEC can be
 satisfied, respectively. Interestingly, we find that DEC is always
 violated, no matter what values the parameters are taken.

 We study the stability properties of this solution. For the little branch,
 the stable point dwells at $z_0=-1.245a^{-1}$, and for the large branch, the stable points
 inhabit at $z_0=-0.3163a^{-1}$ and $z_0=0.9770a^{-1}$.

 We do some primary researches on matching to vacuum solutions.
 Minkowski's and Taub's are studied respectively. Furthermore,  for
 clarifying the physical contents of this solution, we should study
 the weak field approximation to find some thing like Newton
 force among test particles.

  The general form of the plane symmetric
  solution (\ref{general}) with a perfect fluid source must imply
  more rich physical and mathematical properties. It may deserve to study further
  in the future. It should be noted that in the present article we have examined only
  the classical stability under $z$-direction perturbations, though it is the
  most important perturbations of plane symmetric space.
   The other forms of classical perturbations, quantum and semi-classical
   perturbations should be thoroughly investigated further.

 {\bf Acknowledgments.}
 H.Noh was supported by grant No. C00022 from the Korea Research
 Foundation. Z.H. Zhu was supported by
  the National Natural Science Foundation of China
    , under Grant No. 10533010, by Program for New Century Excellent
    Talents in University (NCET) and SRF for ROCS, SEM of China.

 \section{appendix}
  One derives the Einstein tensor of metric (\ref{general}), whose
  components read,
  \bea
  G_t^t=\left[-16b^3c^4du^3+16b^4c^3du^5+12b^3c^4du^3+45b^2c^4u^2v^2
  +3b^7d^2cu^{10}+12b^{7/2}c^{5/2}u^5v+18b^2c^6d^2
  - \right.\n \\ 12b^4c^3du^5+
  45b^3c^5d^2u^2+18b^6c^2d^2u^8+45b^5c^3d^2u^6
  +18bc^5v^2-6b^6\s{bc}d u^{10}v+18b^{9/2}c^{3/2}u^{7}v-
  \n \\
  16b^2c^4e^{2ar}-120(bc)^{7/2}du^4v+6b^5cu^8+6b^{11/2}c^{1/2}u^9v-24b^2c^5du-
  8bc^6/u+24b^{3/2}c^{9/2}v+ \n \\
   8\s{bc}c^5u^{-1}v-12b^{5/2}c^{7/2}u^3v+45b^4c^2u^6v^2+8b^4c^2u^6+24b^5c^2du^7
   +60b^4c^4d^2u^4-90b^{9/2}c^{5/2}dv+ \n \\
   b^{5/2}c^{9/2}du+8b^6cdu^9-8bc^5-6\s{bc}c^6u^{-2}v+18b^5cu^8v^2+3c^6u^{-2}v^2
   +6bc^5-16b^2c^3u^4+ \n \\
   3bc^7d^2u^{-2}-18b^{3/2}c^{9/2}uv-
   6b^{1/2}c^{11/2}u^{-1}v-8b^5cu^8+4b^3c^3u^4-18b^5c^2du^7+6bc^6d/u- \n \\
   16b^4c^2u^6+3b^6u^{10}v^2-36b^{3/2}c^{11/2}v+16b^2c^3\s{bc}u^3v-16b^4c^3u^5v
   -90b^2c^4\s{bc}du^2v-24b^5c^2u^7v+ \n \\
   \left.+8b^2c^4u^2-36b^5c\s{bc}du^8v-8b^5\s{bc}v+60b^3c^3u^4v^2-6b^6cdu^9\right]a^2(bc)^{-1}
   (c+bu^2)^{-4}~,
     \ena
   \bea
   G_z^z=-\left[4bc^6du^{-1}-6b^2c^6d^2-2b^5\s{bc}u^7v+40b^3c^3\s{bc}dv/u-
   8b^3c^3u^2+2\s{bc}c^6u^{-2}v-6b^2c^5d/u- \right. \n \\
   4bc^5+6b^5c^2du^5-15b^4c^2u^4v^2-8b^4c^3du^3-12b^5c^2du^5-6b^4c\s{bc}u^5v
   -4b^5cu^6+\s{bc}c^5vu^{-3}+ \n \\
   2b^6cdu^7+2b^6\s{bc}du^8v+12b^4c\s{bc}u^5v+12b^5c\s{bc}u^6v
   +8b^3c^4du-15b^2c^4v^2-6b^6c^2d^2u^6-\n \\
   6bc^5u^{-2}v^2-b^6u^8v^2-4b^6cdu^7+30b^2c^4\s{bc}dv+4b^4c^3du^3+8b^3c^3u^2
   +4b^5cu^6+6bc^4\s{bc}v/u+ \n \\
   12bc^5\s{bc}u^{-2}v-15b^3c^5d^2+30b^4c^2\s{bc}u^4v+8b^3c^2\s{bc}u^3v
   +4bc^5\s{bc}u^{-2}-2bc^6\s{bc}du^{-3}-\n \\
   c^6u^{-4}v^2-b^7cd^2u^8-4b^3c^4du+4b^5\s{bc}u^7v+12b^2c^5d/u+4b^2c^3\s{bc}
   uv-bc^7d^2u^{-4}-12bc^4\s{bc}v/u+ \n \\
   \left.8b^2c^4-20b^4c^4d^2u^2-20b^3c^3u^2v^2-8b^{3}c^3\s{bc}uv-4c^5\s{bc}u^{-3}v
   -6b^5cu^6v^2\right]a^2(bc)^{-1}
   (c+bu^2)^{-4}~,
   \ena
   where $u=e^{az}$, $v=\ar\left(u\s{\frac{b}{c}}\right)$ and
   \be
   G_{x}^{x}=G_y^y=G_z^z.
   \en
    The Einstein tensor is checked by  software Maple. We see that, though the Einstein tensor is rather messy, it
    reveals a significant property of metric (\ref{general}): it
    describes a solution of plane symmetric sourced by  perfect fluid,
    whose energy momentum is in the form (\ref{em}). Then in coordinate system
    (\ref{general}),
    \bea
    T_t^t=G_{t}^{t},  \\
    T_{x}^{x}=T_y^y=T_z^z=G_{z}^{z}.
    \ena
    We use the convention $8\pi G\equiv 1$ throughout this article.
    So, in the comoving system of the perfect fluid, $U=\partial/\partial t$,
    \bea
    \label{density}
    \rho=-T_t^t,  \\
    p=T_z^z.
    \label{pressure}
    \ena

\end{document}